\documentclass[conference]{IEEEtran}
\usepackage{comment}

\usepackage{cite}
\usepackage{amsmath,amssymb,amsfonts}
\usepackage{algorithmic}
\usepackage{graphicx}
\usepackage{textcomp}

\usepackage{cite}
\usepackage{amsmath,amssymb,amsfonts}
\usepackage{algorithmic}
\usepackage{graphicx}
\usepackage{epstopdf}
\usepackage{subfigure}
 \graphicspath{{\figures}}
\usepackage{textcomp}
\usepackage{xcolor}
\usepackage{verbatim}
\usepackage{makecell}
\usepackage{booktabs}
\usepackage{CJK}
\usepackage{multirow}
\usepackage{tabularx}

\usepackage{algorithm}
\usepackage{algorithmic}
\usepackage{indentfirst}
\usepackage{subcaption}
\makeatletter
\newcommand{\removelatexerror}{\let\@latex@error\@gobble}
\makeatother

\IEEEoverridecommandlockouts   
\begin{document}


\title{Video Semantic Communication with Major Object Extraction and Contextual Video Encoding}

\author{Haopeng Li\IEEEauthorrefmark{1}, Haonan Tong\IEEEauthorrefmark{1}, Sihua Wang\IEEEauthorrefmark{1}, Nuocheng Yang\IEEEauthorrefmark{1}, Zhaohui Yang\IEEEauthorrefmark{2}, and Changchuan Yin\IEEEauthorrefmark{1}\\
\small \IEEEauthorrefmark{1}  
Beijing Laboratory of Advanced Information Network, Beijing University of Posts and Telecommunications, Beijing, China\\
\IEEEauthorrefmark{2}
College of Information Science and Electronic Engineering, Zhejiang University, Hangzhou, China\\
Emails: \{hpli, hntong, sihuawang, yangnuocheng, ccyin\}@bupt.edu.cn, yang\_zhaohui@zju.edu.cn
\vspace{-0.4cm}
\thanks{This work was supported in part by Beijing Natural Science Foundation under Grant L223027, in part by the National Natural Science Foundation of China under Grants 61671086 and 61629101, supported by National Key R\&D Program of China (Grant No.2023YFB2904804), in part by BUPT Excellent Ph.D. Students Foundation under Grant CX2021114, and China Scholarship Council.}
}
\maketitle

\begin{abstract}
This paper studies an end-to-end video semantic communication system for massive communication. 
In the considered system, the transmitter must continuously send the video to the receiver to facilitate character reconstruction in immersive applications, such as interactive video conference. 
However, transmitting the original video information with substantial amounts of data poses a challenge to the limited wireless resources. 
To address this issue, we reduce the amount of data transmitted by making the transmitter extract and send the semantic information from the video, which refines the major object and the correlation of time and space in the video. 
Specifically, we first develop a video semantic communication system based on major object extraction (MOE) and contextual video encoding (CVE) to achieve efficient video transmission. 
Then, we design the MOE and CVE modules with convolutional neural network based  motion estimation, contextual extraction and entropy coding. 
Simulation results show that compared to the traditional coding schemes, the proposed method can reduce the amount of transmitted data by up to 25\% while increasing the peak signal-to-noise ratio (PSNR) of the reconstructed video by up to 14\%. 

\end{abstract}

\vspace{-0.1cm}
\section{Introduction}
The development of B5G/6G communication has promoted the vigorous rise of new broadband interactive applications (e.g. metaverse), which need to collect and transmit massive sensing data in different modals~\cite{yzh}. 
For example, in the interactive application such as immersive video conference, the transmitter needs to send information including the characters, the actions, and the background to the receiver.  
However, the interactive application requires a large amount of data transmission~\cite{videoNumber} and strict transmission performance, such as stringent delay~\cite{gwt} and high quality of experience,  
which brings a huge challenge to the existing wireless communication system with limited resources~\cite{bandwidthConstrain}. 
Therefore, future interactive applications call for efficient video transmission technology, and semantic communication~\cite{thn} is emerging as a high-efficiency transmission technology to improve the broadband interactive application performance. 

Previous works have studied semantic communication techniques for different data types. 
In~\cite{semanticSystemExample}, the authors proposed a deep learning (DL) based semantic communication system to reduce the amount of text information data. 
Furthermore, the work in~\cite{semanticSystemExampleImage} presented the generative adversarial networks (GANs) based image semantic codec to reduce the amount of image transmission data by transmitting semantic information rather than original symbols. 
Extending the single user scenario, the authors in~\cite{multimodal} proposed a DL based multi-user semantic communication system to transmit multi-modal data.  
However, the semantics used for transmitting text and image in~\cite{semanticSystemExample,semanticSystemExampleImage,multimodal} only considered the data properties for one moment, which ignored the semantics from the temporal correlation in the data stream. 
Different from text and image, video semantic extraction is challenging due to the correlations among multiple moments within the video data streams. 

The prior works~\cite{DVC,SemanticCommunicationsFirst,ZhangPing} have studied efficient video semantic communication systems. 
In~\cite{DVC}, the authors proposed an end-to-end DL based video communication model that obtained the motion information among video frames to encode the video. 
The work in~\cite{SemanticCommunicationsFirst} presented a semantic video conferencing~(SVC) model based on face key points transmission to express the face motions. 
The authors in~\cite{ZhangPing} proposed a scheme that collected the strong video temporal correlation provided by the feature domain context under the deep video semantic transmission~(DVST) model to achieve more efficient video transmission. 
However, for interactive applications such as video conferences, the static background information is often redundant and unnecessary to be transmitted in each frame. 
Therefore, removing redundant background information according to the need of reconstructing content in video transmission can reduce the amount of transmission data and decrease transmission delay, which has not been considered in~\cite{DVC,SemanticCommunicationsFirst,ZhangPing}. 


To address the above issue, in this paper, \textit{we propose a video semantic communication system, which extracts the major object of the video for character reconstruction at the receiver}. 
\begin{figure*}
    \centering
    \includegraphics[width=1\linewidth]{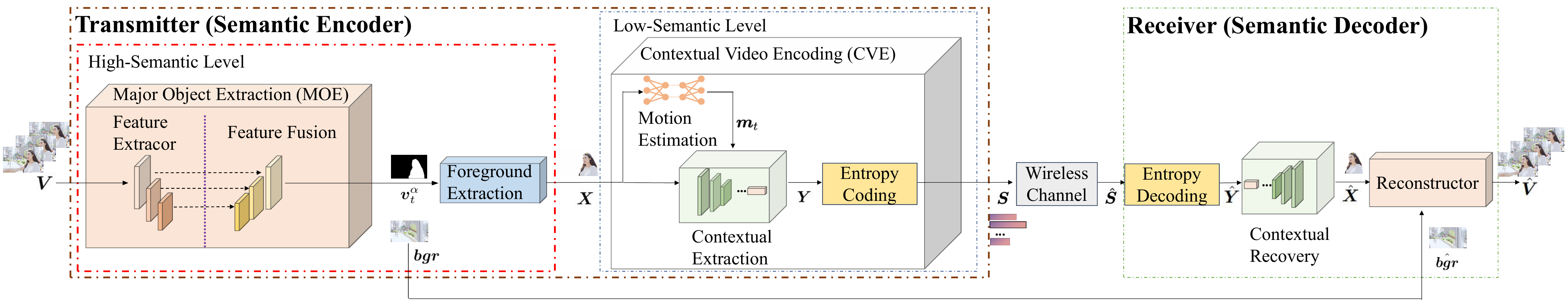}
    \setlength{\abovecaptionskip}{0cm}
    \caption{The overall flowchart of the proposed MOE-CVE scheme.}\label{fig:sys}
    \vspace{-0.6cm}
\end{figure*}
The main contributions of this paper are as follows:

(1) We develop a video semantic communication system through major object extraction (MOE)-contextual video encoding (CVE). 
In the considered system, the transmitter extracts the semantic information of the major object in the video, through removing the redundant static background, and then sends the semantic information of the major object to the receiver. 
The receiver decodes the semantic information and synthesizes the background. 

(2) We propose an efficient video semantic extraction method using hybrid high-level and low-level semantic information, where the high-level semantic is the major object information, and the low-level semantic is the spatial structure in the video frame. 
To achieve the above goal, we design the MOE module to extract the major object, and the CVE module to semantically encode video at a low-semantic level, in which context based spatial structure is extracted and encoded by entropy coding. 

(3) Simulation results demonstrate that the proposed MOE-CVE scheme can reduce the amount of transmitted data by up to 25\% while increasing peak signal-to-noise ratio (PSNR) by up to 14\%, compared to the traditional coding schemes. 

The rest of this paper is organized as follows. 
Section II introduces the system model. 
Section III provides detailed descriptions of the proposed network architectures.
Simulation results are presented in Section IV. 
Conclusion is drawn in Section V. 
\vspace{-0.3cm}
\section{System Model}
We consider a video semantic communication system to support interactive video conference, which includes a transmitter~\cite{uav} and a receiver as shown in Fig.~\ref{fig:sys}. 
The receiver needs to accurately reconstruct the character information, while the static background information does not need to be updated in time, and only limited wireless bandwidth is available to support the massive communication.
To this end, the system requires efficient video transmission techniques by extracting and transmitting video semantic information. 
To achieve this goal, we consider a hybrid high-level and low-level semantic extraction by removing the redundant static video background information, which mainly includes three steps. 
First, we extract the major object of the video at a high-semantic level, where video frames are divided into foreground and background, the static background is transmitted only once, and the foreground needs to be updated in real time.  
Second, the foreground is further encoded through temporal and spatial correlations at a low-semantic level. 
Finally, given the transmitted foreground semantic feature, the receiver reconstructs the video frames by combining the character in the foreground and the one-frame static background into a complete video. 

Next, we first introduce the MOE at the high-semantic level. 
Then, we present our CVE at the low-semantic level.  
Subsequently, we introduce our channel model. 
Finally, we introduce the method for decoding and reconstructing video.
\vspace{-0.1cm}
\subsection{\textit{Major Object Extraction}}
Denote the raw video data sensed at the transmitter by $\boldsymbol{V}=\left \{ \boldsymbol{v}_{1}, \boldsymbol{v}_{2}, \cdots\cdots, \boldsymbol{v}_{T} \right \}$, where $\boldsymbol{v}_{t}\in \mathbb{R} ^{H\times W\times C}$ is the $t$-th video frame. 
$H$ and $W$ are the height and width of the video frame, respectively.
$C$ is the number of channels. 
The MOE module is used to extract the major object from $\boldsymbol{V}$, and the major object is the foreground component. 
Denote the foreground estimation tensor by $\boldsymbol{v}_{t}^{\mathrm{\alpha}}\in \mathbb{R} ^{H\times W\times 1}$, and $\boldsymbol{v}_{t}^{\mathrm{\alpha}}$ can be given by
\begin{align}
\boldsymbol{v}_{t}^{\mathrm{\alpha}} = f_{MOE} ( \boldsymbol{v}_{t} ),\label{alpha}
\end{align}
where $f_{MOE}(\cdot )$ is the function of the MOE module. 

Subsequently, we employ the foreground extraction to obtain the foreground series $\boldsymbol{X}=\left \{ \boldsymbol{x}_{1}, \boldsymbol{x}_{2}, \cdots\cdots, \boldsymbol{x}_{T} \right \}$ with $\boldsymbol{v}_{t}^{\mathrm{\alpha}}$, where the $t$-th frame $\boldsymbol{x}_{t}\in \mathbb{R} ^{H\times W\times C}$ 
undergoes background pixel removal, and can be given by 
\begin{align}
\boldsymbol{x}_{t}= \boldsymbol{v}_{t}^{\mathrm{\alpha}}\odot\boldsymbol{v}_{t}+(\boldsymbol{1}-\boldsymbol{v}_{t}^{\mathrm{\alpha}})\odot\boldsymbol{b}  \label{compose},
\end{align}
where $\boldsymbol{b}$ is a white pixel tensor to represent the blank background,  
and $\odot$ is the Hadamard product. 
Given $\boldsymbol{v}_{t}$ and $\boldsymbol{x}_{t}$, the background can be extracted with 
$\boldsymbol{bgr} = \boldsymbol{v}_{t}-\boldsymbol{x}_{t}$, where the background $\boldsymbol{bgr}\in \mathbb{R} ^{H\times W\times C}$ can be further completed by image generation techniques. 
\vspace{-0.1cm}
\subsection{\textit{Contextual Video Encoding}}
To take advantage of the temporal and spatial correlations of the foreground to further reduce the amount of data required for transmission, we transform $\boldsymbol{x}_{t}$ from the pixel domain to the semantic feature $\boldsymbol{y}_{t}$ in the latent domain and then encode $\boldsymbol{y}_{t}$ to the variable-length code words $\boldsymbol{s}_{t}$ for transmission.
In the CVE module, we first use the motion estimation network to learn the motion vector $\boldsymbol{m}_{t}\in \mathbb{R} ^{H\times W\times (C-1)}$ between the current frame $\boldsymbol{x}_{t}$ and the previous frame $\boldsymbol{x}_{t-1}$, which is given by
\vspace{-0.3cm}
\begin{align}
    \boldsymbol{m}_{t} =f_{Est} (\boldsymbol{x}_{t},\boldsymbol{x}_{t-1}),
\end{align}
where $f_{Est}(\cdot)$ denotes the motion estimation function.  
Since neural network can effectively refine the video features in the latent domain~\cite{nonlinear}, given the motion vector $\boldsymbol{m}_{t}$, we can refine the original video foreground series $\boldsymbol{X}$ into the semantic feature series $\boldsymbol{Y}=\left \{ \boldsymbol{y}_{1}, \boldsymbol{y}_{2} ,\cdots\cdots, \boldsymbol{y}_{T} \right \}$, which is in a latent domain representation with smaller data size as shown in Fig.~\ref{fig:sys}. 
The relationship between $\boldsymbol{y}_{t}$ and $\boldsymbol{x}_{t}$ can be given by
\begin{align}
\boldsymbol{y}_{t}= f_{CE} \left (\boldsymbol{x}_{t} , \boldsymbol{m}_{t}\right), \label{latent transform}
\end{align}
where $f_{CE}(\cdot)$ denotes the contextual extraction function, and $\boldsymbol{y}_{t}\in \mathbb{R} ^{\frac{H}{16} \times \frac{W}{16}\times 96}$ is the semantic feature of a single frame. 
Finally, $\boldsymbol{y}_{t}$ is encoded into code words streaming $\boldsymbol{S}=\left \{ \boldsymbol{s}_{1}, \boldsymbol{s}_{2} ,\cdots\cdots, \boldsymbol{s}_{T} \right \}$ by entropy coding algorithm to transmit through wireless channel, which can be given by 
\begin{align}
\boldsymbol{s}_{t}= f_{en} \left (\boldsymbol{y}_{t}\right),\label{contextual encoder}
\end{align}
where $f_{en}(\cdot)$ denotes the function of encoding, and $\boldsymbol{s}_{t}\in \mathbb{R} ^{k_{t}}$ is the variable-length code words with $k_{t}$ being the length of $\boldsymbol{s}_{t}$. 
We use $R=\frac{1}{T}\sum_{t=1}^{T} \frac{k_{t} }{H\times W\times C}$ as the channel bandwidth ratio (CBR)~\cite{ZhangPing} to describe the average coding rate of $\boldsymbol{X}$.
\begin{figure}[t]
    \centering
    \includegraphics[width=0.9\linewidth]{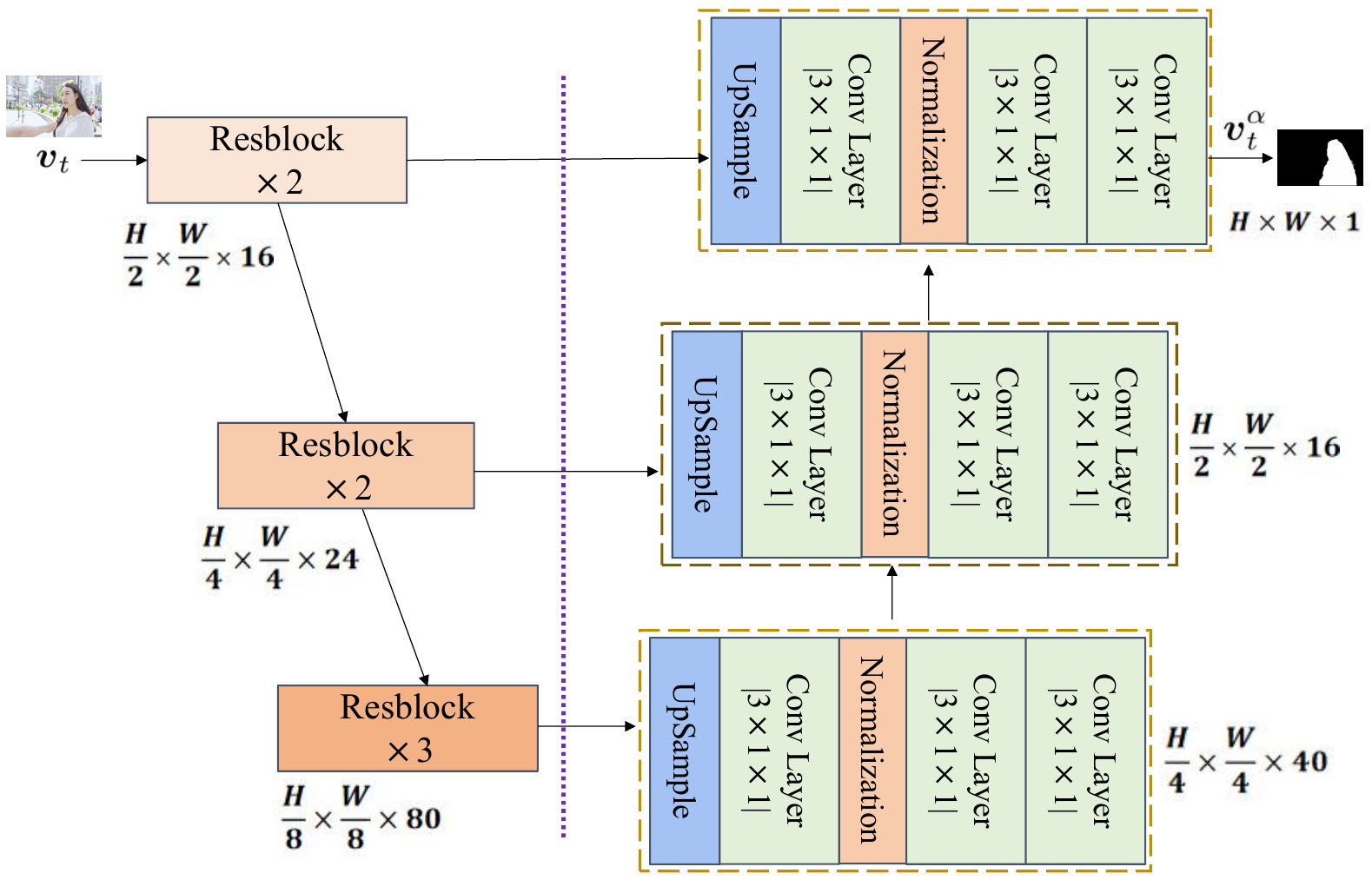}
    \setlength{\abovecaptionskip}{0cm}
    \caption{The structure of MOE, where the Conv Layer means the convolutional layer with $|kernel\times stride\times padding|$, and the Resblock means the residual block.}\label{MOE}
    \vspace{-0.6cm}
\end{figure}
\vspace{-0.1cm}
\subsection{\textit{Wireless Channel}}
When transmitted over a wireless channel, encoded code words streaming $\boldsymbol{S}$ suffers transmission impairments that include distortion and noise. 
Assume that the video transmission uses a single end-to-end wireless link, the received signal can be given by
\vspace{-0.4cm}
\begin{align}
\hat{\boldsymbol{S}}  = h \boldsymbol{S}  + \boldsymbol{\sigma},
\end{align}
where $\hat{\boldsymbol{S}}$ is the received semantic feature code words streaming with transmission impairment, $h$ is the fading channel coefficient, and $\boldsymbol{\sigma} \sim \mathit{\mathcal{N} } \left ( 0,\sigma ^{2}\boldsymbol{I} \right )$ denotes Gaussian channel noise with $\sigma ^{2}$ being noise variance and $\boldsymbol{I}$ being identity matrix.
\vspace{-0.1cm}
\subsection{\textit{Receiver}}
The receiver includes the entropy decoding, the contextual recovery, and the reconstructor modules.
The entropy decoding is used to decode the received code words streaming $\hat{\boldsymbol{S}}$ to the semantic feature $\hat{\boldsymbol{Y}}$, which is given by
\begin{align}
\hat{\boldsymbol{Y}} = f_{de}( \hat{\boldsymbol{S}}). \label{contextualdecoder}
\end{align}
Then the contextual recovery is used to obtain $\hat{\boldsymbol{X}}$ from semantic feature $\hat{\boldsymbol{Y}}$, given by
\vspace{-0.2cm}
\begin{align}
\hat{\boldsymbol{X}} = f_{CR}( \hat{\boldsymbol{Y}}).\label{latent inversion}
\end{align}

The reconstructor is used to combine the static background that is directly provided at the receiver.  
In the considered system, the static background can be reused for multiple video frames. 
As a result, we transmit one static background frame and use it as the background of the reconstructed video $\hat{\boldsymbol{bgr}}$.  
Finally, we combine the video frames back to reconstructed video $\hat{\boldsymbol{V}}=\left \{ \hat{\boldsymbol{v}}_{1}, \hat{\boldsymbol{v}}_{2} ,\cdots\cdots, \hat{\boldsymbol{v}}_{T}\right \}$, as shown in Fig.~\ref{fig:sys}. 

\section{Network Architectures}
\vspace{-0.1cm}
In this section, we design the system module to implement the proposed system jointly considering accurate semantic extraction and various channel conditions. 
In particular, we first extract the high-level semantic in each single frame via neural network based MOE in semantic encoder. 
Then, we extract the low-level semantic through CVE in semantic encoder, considering correlations between frames, and adaptive various length coding, thus reducing the transmitted data amount.
Finally, the semantic decoder and the whole training algorithm are illustrated. 
\begin{figure}[t]
    \centering
    \includegraphics[width=1\linewidth]{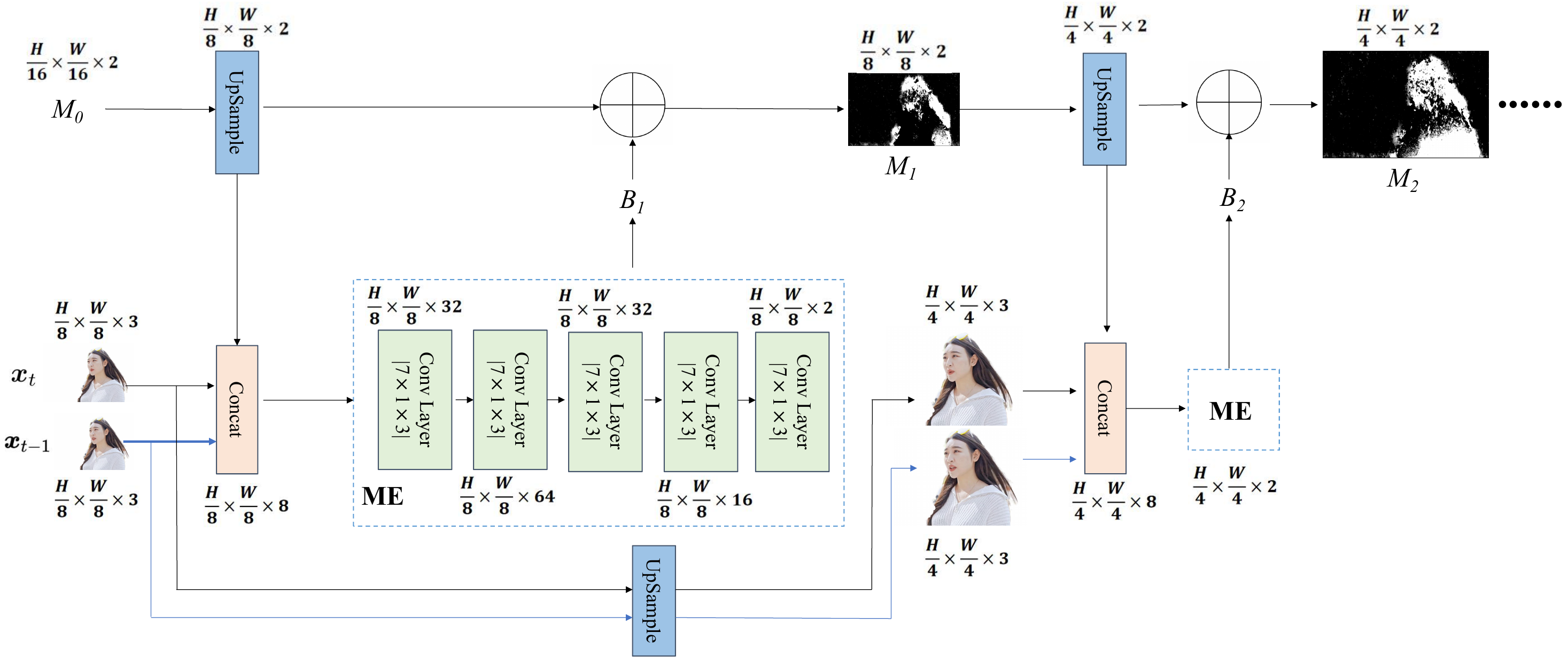}
    \setlength{\abovecaptionskip}{0cm}
    \caption{Optical flow estimation network.}\label{optical}
    \vspace{-0.4cm}
\end{figure}
 \vspace{-0.1cm}
\subsection{\textit{MOE in Semantic Encoder}} 
The MOE module is deployed to perform character extraction from the input video frames. 
It first recognizes the major object, and then removes the static background. 
In this way, we consider the object-level content as the high-level semantics. 
To accurately recognize the major object, we extract features at different scales which include object shapes and locations. 
We use MOE module to generate $\boldsymbol{v}_{t}^{\mathrm{\alpha}}$ in (\ref{alpha}). 
As shown in Fig.~\ref{MOE}, we use MobileNetV3-Large~\cite{MOE} as the backbone of MOE module, which is composed of a feature extractor and a feature fusion. 
The feature extractor consists of residual blocks to extract features at $\frac{1}{2}$, $\frac{1}{4}$, $\frac{1}{8}$ scales of the video frame, respectively, to accurately locate video major object. 
The feature fusion part fuses features with different scales via convolutional layers to generate $\boldsymbol{v}_{t}^{\mathrm{\alpha}}$, as shown in Fig.~\ref{MOE}. 

To improve the accuracy of foreground estimation, the loss function of MOE module to output $\boldsymbol{v}_{t}^{\mathrm{\alpha}}$ is given by
\vspace{-0.1cm}
\begin{align}
\mathcal{L}_{MOE}  = 
\left \| \boldsymbol{v}_{t}^{\mathrm{\alpha}} -\boldsymbol{v}_{t}^{\mathrm{\alpha*}} \right \| _{1}
+\mathcal{L} ^{\alpha }_{lap} +5
\left \| \frac{\mathrm{d} \boldsymbol{v}_{t}^{\mathrm{\alpha}} }{\mathrm{d} t} 
-\frac{\mathrm{d} \boldsymbol{v}_{t}^{\mathrm{\alpha*}}}{\mathrm{d} t} \right \|_{2},\label{loss_MOE}
\end{align}
where $\boldsymbol{v}_{t}^{\mathrm{\alpha*}}$ is the ground-truth of $\boldsymbol{v}_{t}^{\mathrm{\alpha}}$, 
and 
\vspace{-0.1cm}
\begin{align}
\mathcal{L} ^{\alpha }_{lap}  = \sum_{i = 1}^{5} \frac{2^{i-1} }{5} 
\left \| \mathit{L} _{pyr} ^{i} (\boldsymbol{v}_{t}^{\mathrm{\alpha}})-
\mathit{L} _{pyr} ^{i} (\boldsymbol{v}_{t}^{\mathrm{\alpha*}}) \right \|_{1}
\end{align}
is the Laplacian loss~\cite{loss-moe-lap} with $\mathit{L} _{pyr} ^{i} (\cdot )$ being the output of the $i$-th pyramid layer~\cite{loss-moe-lap}. 
\vspace{-0.1cm}
\subsection{\textit{CVE in Semantic Encoder}} 
In CVE, we first use motion vector $\boldsymbol{m}_{t}$ to refine the semantic feature $\boldsymbol{y}_{t}$ from $\boldsymbol{x}_{t}$ with contextual extraction.  
Then $\boldsymbol{y}_{t}$ is encoded as variable length code words $\boldsymbol{s}_{t}$, based on the entropy estimation of $\boldsymbol{y}_{t}$. 
Next, we will introduce the motion estimation, the contextual extraction, and the entropy encoding in turn.
\subsubsection{\textit{Motion Estimation}}
Since the foreground has strong temporal and spatial correlations between successive video frames, the low-level semantic features (temporal and spatial correlations) are further extracted by CVE module to reduce transmission overhead.
To obtain  the contextual information $\boldsymbol{z}_{t}\in \mathbb{R} ^{H\times W\times 64}$ which is the combination of temporal and spatial correlations between successive video frames, we first use convolutional neural network (CNN) to estimate motion vector $\boldsymbol{m}_{t}$. 
Compared with traditional motion estimation (ME) methods, neural network can integrate multi-dimensional information such as color, texture, and depth from videos to improve the accuracy of motion estimation, so we estimate the motion vector $\boldsymbol{m}_{t}$ with CNN.
ME is position change prediction of the pixel blocks.
In the ME process, we use the optical flow estimation network~\cite{opticalEstimation} to estimate the motion vector $\boldsymbol{m}_{t}$ between the previous frame $\boldsymbol{x}_{t-1}$ and the current frame $\boldsymbol{x}_{t}$. 
Concretely, the optical flow estimation network consists of four ME modules to form four pyramid levels~\cite{opticalEstimation}. 
As shown in Fig.~\ref{optical}, the optical flow estimation uses a coarse-to-fine spatial pyramid structure to learn residual flow $B_{k}\in \mathbb{R} ^{\frac{H}{2^{4-k}} \times \frac{W}{2^{4-k}}\times 2}$ at each pyramid level, and the optical flow $M_{k}\in \mathbb{R} ^{\frac{H}{2^{4-k}} \times \frac{W}{2^{4-k}}\times 2}$ at the $k$-th pyramid level is given by
\begin{align}
M_{k}  = u(M_{k-1})+B_{k}, ~ k=1,2,3,4,
\end{align}
where $u(\cdot )$ is the upsampling operation, $B_{k}$ is the residual flow output by the ME module, and
$M_{0}\in \mathbb{R} ^{\frac{H}{16} \times \frac{W}{16}\times 2}$ is a zero tensor as the initial value.
We use the ME module to calculate the residual optical flow $B_{k}$ of the current level ($k$-th level). 
The residual optical flow $B_{k}$ is input successively to rectify the optical flow $M_{k}$ at each level.
ME is the last step of optical flow estimation, and finally output the last optical flow as the motion vector, given by $\boldsymbol{m}_{t}=M_{4}$.
\subsubsection{\textit{Contextual Extraction}}
\begin{figure}[t]
    \centering
     \includegraphics[width=0.9\linewidth]{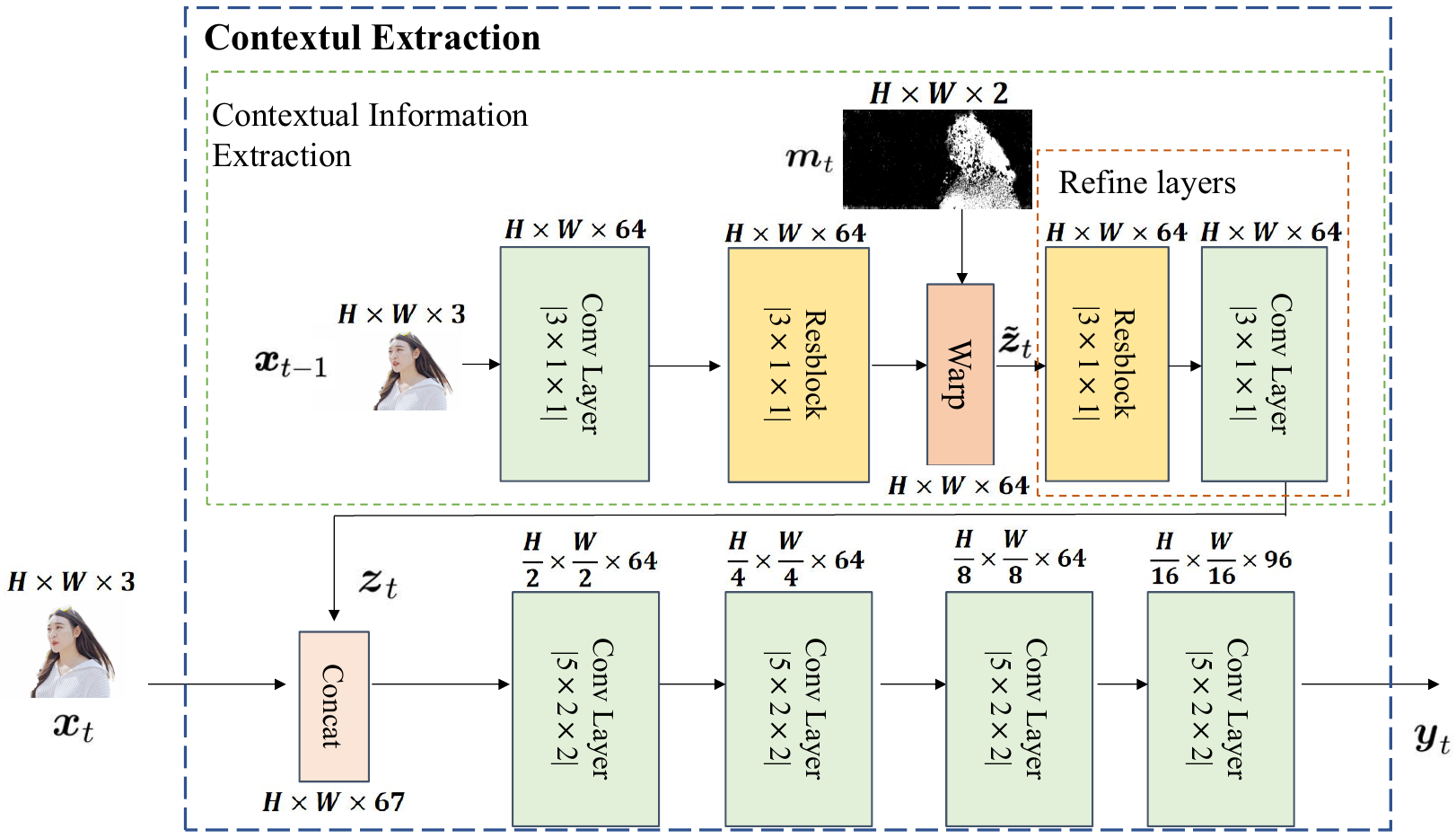}
    \setlength{\abovecaptionskip}{0cm}
    \caption{The structure of contextual extraction.}\label{contextual_extraction}
    \vspace{-0.6cm}
\end{figure}
As shown in Fig.~\ref{contextual_extraction}, we use the obtained $\boldsymbol{m}_{t}$ and the previous frame to extract the initial contextual information $\tilde{\boldsymbol{z}}_{t}\in \mathbb{R} ^{H \times W\times 64}$ with convolutional layers and residual blocks.
Here, to compensate for the discontinuity~\cite{ZhangPing}, we use refine layers to refine the final contextual information $\boldsymbol{z}_{t}\in \mathbb{R} ^{H \times W\times 64}$ which contains temporal and spatial correlations.
Then we use $\boldsymbol{z}_{t}$ to extract the semantic feature $\boldsymbol{y}_{t}$ with CNN layers, where $\boldsymbol{y}_{t}$ is the mapping feature of $\boldsymbol{x}_{t}$ in the latent domain. 
Instead of directly encoding the video frame in the pixel domain in traditional methods, we encode semantic feature $\boldsymbol{y}_{t}$ taking into account both the temporal and spatial correlations in the regions across video frames, which is more conducive to the subsequent allocation of coding weights. 
\subsubsection{\textit{Entropy Coding}}
Given $\boldsymbol{y}_{t}$, we encode the semantic feature $\boldsymbol{y}_{t}$ into the variable-length code words $\boldsymbol{s}_{t}$ with entropy coding. 
\begin{figure}
    \centering
     \includegraphics[width=0.9\linewidth]{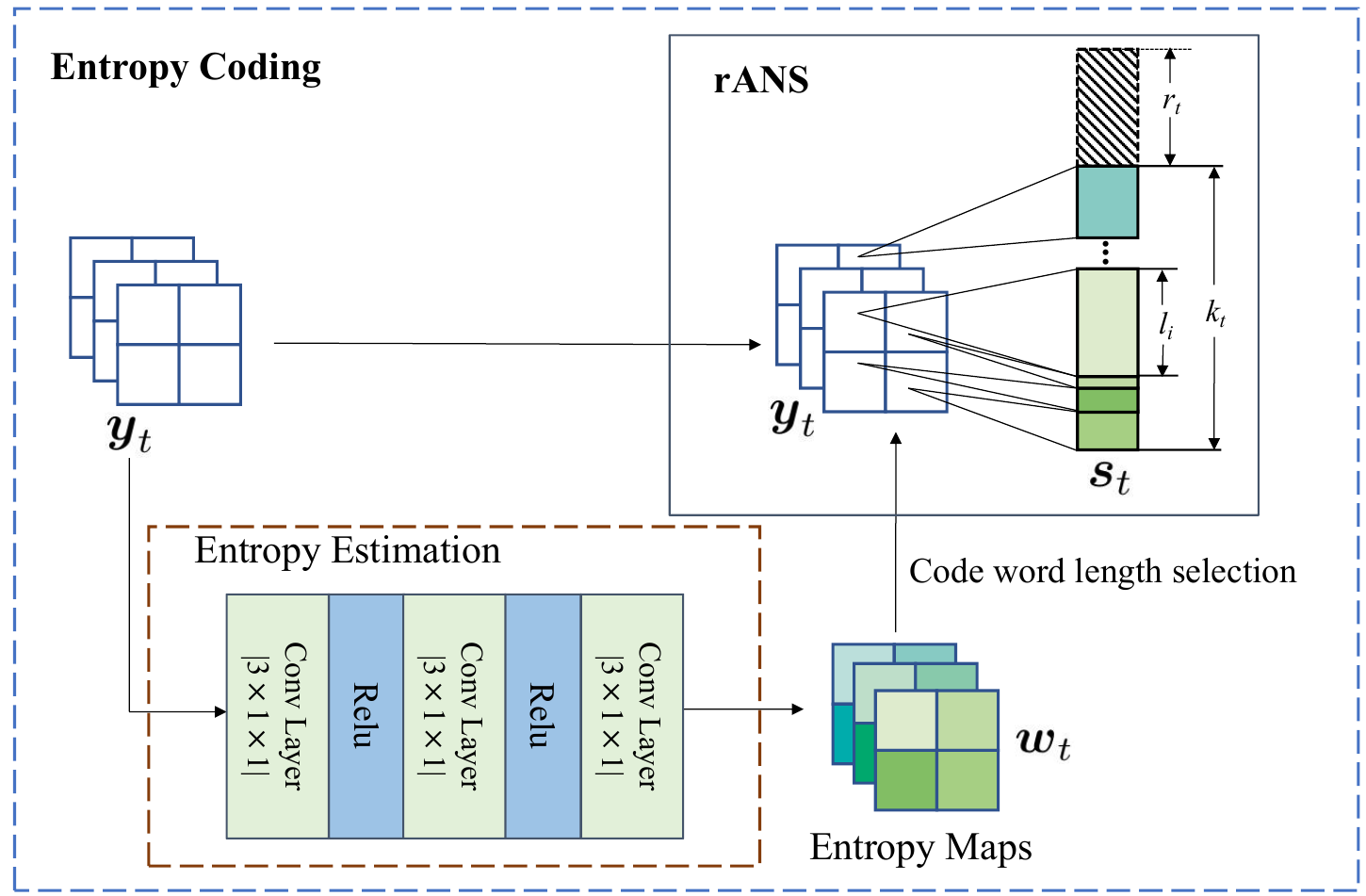}
    \setlength{\abovecaptionskip}{0cm}
    \caption{The structure of entropy coding.}\label{entropyEncoding}
    \vspace{-0.6cm}
\end{figure}
As shown in Fig.~\ref{entropyEncoding}, the entropy coding consists of the entropy estimation and a range asymmetric numeral system (rANS) arithmetic encoder~\cite{rANS}. 
 The entropy estimation is composed of the convolutional layers and the relu layers, which assigns weights to the elements of semantic feature $\boldsymbol{y}_{t}$ and generates a weight vector $\boldsymbol{w}_{t}\in \mathbb{R} ^{\frac{H}{16} \times \frac{W}{16}\times 96}$~\cite{mdvsc}. 
Each element in $\boldsymbol{w}_{t}$ represents a weight that maps the elements in $\boldsymbol{y}_{t}$ to the code words space $\Omega \in \mathbb{R} ^{64 \times 103}$.
In detail, each element in $\Omega $ follows the Laplacian distribution. 
The element in $\boldsymbol{w}_{t}$ determines the mapped position in $\Omega $ of the element in $\boldsymbol{y}_{t}$, thus deciding the code word length ${k}_{t}$. 
In this way, a greater estimated entropy in $\boldsymbol{y}_{t}$ is mapped by $\boldsymbol{w}_{t}$ to a longer code length in $\Omega $.
The code word length ${k}_{t}$ of $\boldsymbol{s}_{t}$ is the sum length of the code words of all the elements in $\boldsymbol{y}_{t}$, which is given by $k_{t} =\sum_{i=1}^{\frac{H}{16} \times \frac{W}{16} \times 96} l_{i} $, where $l_{i}$ is the code word length of $i$-th element in $\boldsymbol{y}_{t}$. 
Then the code words streaming $\boldsymbol{S}$ is transmitted to the wireless channel. 
As shown in Fig.~\ref{entropyEncoding}, we only encode characters for transmission, saving the code word length ${r}_{t}$ required for transmitting the redundant background. 

\vspace{-0.2cm}
\subsection{\textit{Semantic Decoder}}
We first decode $\hat{\boldsymbol{s}}_{t}$ to $\hat{\boldsymbol{y}}_{t}$ by arithmetic decoding in rANS as shown in Fig.~\ref{fig:sys}, which is the inverse process of entropy coding. 
Then we use the contextual recovery which consists of the convolutional layers, the upsample layers, and the residual blocks to recover $\hat{\boldsymbol{y}}_{t}$ to $\hat{\boldsymbol{x}}_{t}$. 
The convolutional layers and the residual blocks are used to reconstruct the video frames from features, 
and the upsample layers are used to gradually recover to the same size as the original video. 
Finally the reconstructor combines the final video $\hat{\boldsymbol{V}}$ by $\hat{\boldsymbol{v}}_{t}= \hat{\boldsymbol{x}}_{t} \odot \hat{\boldsymbol{x}}_{t}^{\mathrm{\alpha}}+(1-\hat{\boldsymbol{x}}_{t}^{\mathrm{\alpha}}) \odot\hat{\boldsymbol{bgr}}$ which is similar to~(\ref{compose}). 
The semantic decoder is trained offline end-to-end with the CVE module. 
In order to control the trade-off between the distortion $d_{t}$ and the bit cost $k_{t}$, the loss of the semantic decoder is
\vspace{-0.2cm}
\begin{align}
\mathcal{L}_{CVE} =k_{t} +\lambda \cdot d_{t}\label{loss-CVE},
\end{align}
where $\lambda$ is the coefficient to control the trade-off between the distortion $d_{t}$ and the bit cost $k_{t}$, $d_{t}$ is peak signal-to-noise ratio (PSNR) of $\hat{\boldsymbol{v}}_{t}$ and $\boldsymbol{v}_{t}$. 
The PSNR can be given by
\vspace{-0.2cm}
\begin{align}
PSNR(\hat{\boldsymbol{v}}_{t},\boldsymbol{v}_{t})= 10\log _{10}(\frac{1}{MSE(\hat{\boldsymbol{v}}_{t},\boldsymbol{v}_{t})}),
\end{align}
where $MSE(\hat{\boldsymbol{v}}_{t},\boldsymbol{v}_{t})=
\frac{1}{mn}\sum_{i=1}^{m}\sum_{j=1}^{n}
[\hat{\boldsymbol{v}}_{t}(i,j)-\boldsymbol{v}_{t}(i,j)]$, with $m$ and $n$ respectively being the number of pixels horizontally and vertically. 

\subsection{\textit{MOE-CVE Training Algorithm}}
 \begin{algorithm}[t]
    \caption{Proposed MOE-CVE Algorithm}
    \label{MOE-CVE Algorithm}
    \begin{algorithmic}[1] 
        \STATE \textbf{Initialize} the parameters of MOE ($\theta$) and CVE ($\omega$).
        \STATE train MOE with (\ref{loss_MOE}): 
        \FOR{each MOE training step}
            \STATE $g_{t}\longleftarrow \bigtriangledown _{\theta } \mathcal{L} _{MOE} (\theta _{t})  \hfill \triangleright \text{calculate gradient}$
            \STATE $e_{t} \longleftarrow \beta _{1} e_{t-1} +(1-\beta _{1})g_{t}   \hfill \triangleright \text{first moment estimation}$
            \STATE $s_{t} \longleftarrow \beta _{2} s_{t-1} +(1-\beta_{2})g_{t}^{2}  \hfill \triangleright \text{second moment estimation} $
            \STATE $\hat{e_{t}} \longleftarrow \frac{e_{t}}{1-\beta _{1} ^{t} } \hfill \triangleright \text{deviation correction} $
            \STATE $\hat{s_{t}}  \longleftarrow \frac{s_{t}}{1-\beta _{2} ^{t}}  \hfill \triangleright \text{deviation correction}$
            \STATE $\theta _{t+1}\longleftarrow \theta _{t}-\frac{\xi }{\sqrt{\hat{s}_{t}}+\epsilon } \hat{e} _{t}  \hfill \triangleright \text{update parameter}$
        \ENDFOR
        \STATE train CVE with (\ref{loss-CVE}):
        \FOR{each CVE training step}
            \STATE $\omega  _{t+1}\longleftarrow \omega  _{t}-\eta \bigtriangledown _{\omega} \mathcal{L}_{CVE}(\omega  _{t})$
        \ENDFOR
        \STATE Receiver reconstructs the video frames.  
        \STATE \textbf{Output:} The reconstructed video $\hat{\boldsymbol{V}}$.
        \end{algorithmic}
\end{algorithm}

 In the proposed system, each module is independent, and we train the MOE module with parameter $\theta$ and the CVE module with parameter $\omega$ separately. 
 In MOE training, we use Adam optimizer, where $g_{t}$ is the gradient of $\mathcal{L} _{MOE}$; $e_{t}$ is the first-moment estimation; $s_{t}$ is the second-moment estimation; $\epsilon$ is $10^{-8}$. 
 The learning rate $\xi$ and $\eta$ for both of MOE and CVE modules is set to $10^{-4}$. 
 After the semantic feature $\boldsymbol{Y}$ is transmitted through the AWGN channel, the receiver uses the received noisy feature $\hat{\boldsymbol{Y}}$ for training. 
 The entropy decoding and the contextual recovery are simultaneous training with CVE in an end-to-end method.
 The overall training process is summarized in Algorithm.~\ref{MOE-CVE Algorithm}.

 \section{Simulation and Performance Analysis}

In our simulations, a video semantic communication system consisting of a transmitter and a receiver is considered. 
Details of all network layers are shown in Figs.\ref{MOE}-\ref{entropyEncoding}.
The MOE model is trained with video dataset VideoMatte240K~\cite{MOEdataset}. 
The dataset provides 484 video clips, and we divide the dataset into 475/4/5 clips for train/val/test splits. 
The CVE module uses Vimeo-90k septuplet dataset~\cite{DCVCtrain} as the training data.
The testing data uses the character dataset provided by~\cite{MOE} which contains a variety of character vide frames with $H$=1920, $W$=1024 and $C$=3. 
We train 4 models with different $\lambda \in \left \{ 256,  512, 1024, 2048 \right \}$ with fading channel with $h$=0.9 and SNR=15dB AWGN.
For comparison purposes, we experiment three baselines:
a) Advanced video coding (H.264) + low-density parity-check (LDPC): H.264 for source coding combined with 1/2 rate LDPC code for channel coding, 
b) High-efficiency video coding (H.265) + LDPC and
c) Deep contextual video compression (DCVC) + LDPC provide the reference here is good if possible.

\begin{figure}[t]
    \centering
    \includegraphics[width=0.8\linewidth, height=0.6\linewidth]{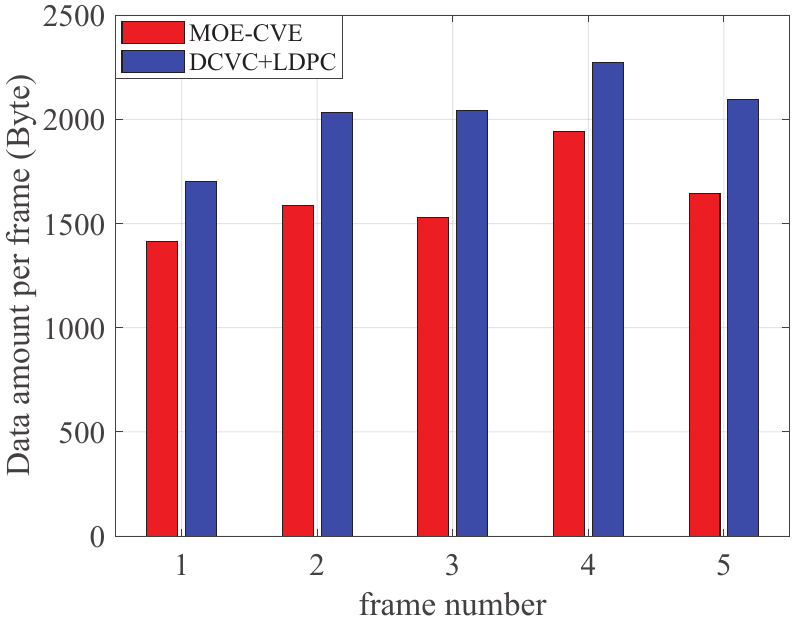}
    \setlength{\abovecaptionskip}{0cm}
    \caption{Data amount of each video frame.}\label{fig:video amount}
    \vspace{-0.3cm}
\end{figure}

In Fig.~\ref{fig:video amount}, we show the transmitted data amount of different video frames in our method and DCVC+LDPC.  
From Fig.~\ref{fig:video amount}, we can observe that both methods use a small amount of data to transmit the original video frame, while our method reduces more data amount. 
From Fig.~\ref{fig:video amount}, we see that our method reduces the data amount of each frame of video by nearly 25\%, compared to DCVC+LDPC.  
This is due to the fact that MOE removes the redundant background information and focuses on the major object. 
Fig.~\ref{fig:video amount} demonstrates that the proposed MOE-CVE method significantly reduces the amount of transmitted data.

\begin{figure}[t]
    \centering
    \includegraphics[width=0.8\linewidth, height=0.6\linewidth]{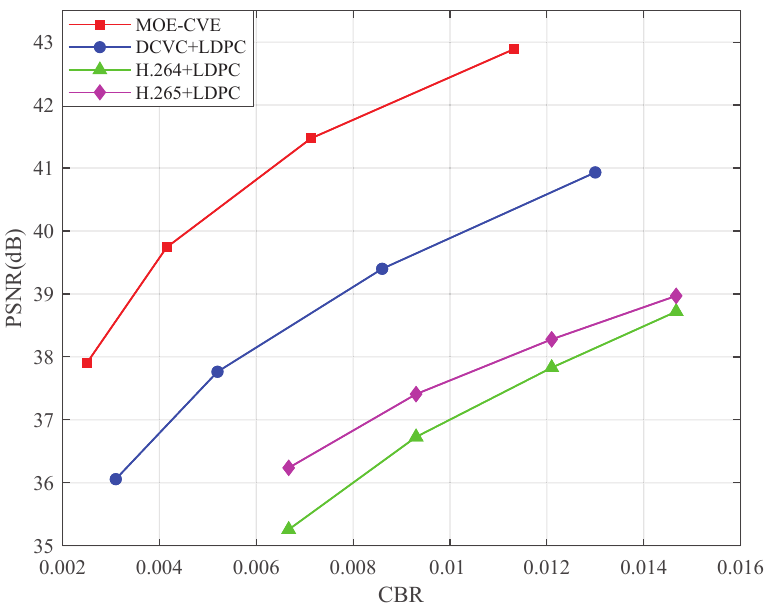}
    \setlength{\abovecaptionskip}{0cm}
    \caption{PSNR of different algorithms under different CBR at SNR = 15dB.}\label{fig:psnr}
    \vspace{-0.7cm}
\end{figure}
In Fig.~\ref{fig:psnr}, we show how the PSNR changes as the CBR varies. 
From Fig.\ref{fig:psnr} we can observe that, as the transmission CBR increases, the transmission PSNR increases as expected, while our MOE-CVE increases more fastly. 
From Fig.\ref{fig:psnr}
We can also observe that, when the CBR is 0.012, the PSNR of our method is 14\% higher than H.264+LDPC, and 12\% higher than H.265+LDPC. 
Moreover, compared to DCVC+LDPC, our method also achieves a improvement in PSNR. 
The improvement is due to the fact that the other algorithms encode all the details of the video frame while our MOE-CVE concentrates on the semantic features of the major object thus achieving better reconstruction performance under the same CBR. 
Fig.~\ref{fig:psnr} demonstrates that the proposed method can improve the video reconstruction accuracy.

\begin{figure}[t]
    \centering
    \includegraphics[width=0.8\linewidth, height=0.6\linewidth]{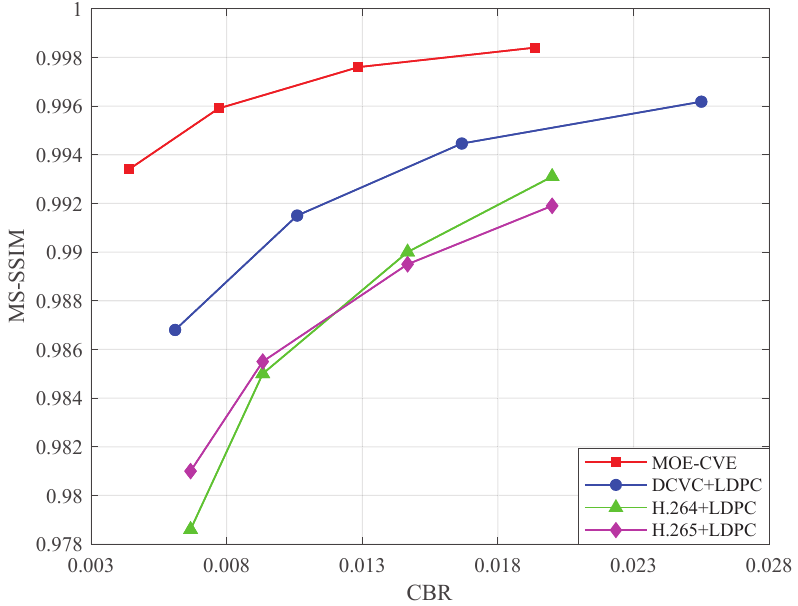}
    \setlength{\abovecaptionskip}{0cm}
    \caption{MS-SSIM of different algorithms under different CBR at SNR = 15dB.}\label{fig:msssim-p}
    \vspace{-0.5cm}
\end{figure}
In Fig.~\ref{fig:msssim-p}, we show how the Multi-scale structural similarity (MS-SSIM)~\cite{mdvsc} changes as the CBR varies. 
From Fig.~\ref{fig:msssim-p} we observe that as the CBR increases, the transmission MS-SSIM increases as expected. 
From Fig.~\ref{fig:msssim-p} we can also observe that our method increases transmission MS-SSIM by nearly 0.014, compared to H.264+LDPC. 
Besides, the MS-SSIM of our method varies flatter than that of the other methods. 
This is because our method removes the statical background, and encodes taking into the correlations, thus improving the MS-SSIM significantly. 

\begin{figure}[t]
    \centering
    \includegraphics[width=0.8\linewidth, height=0.6\linewidth]{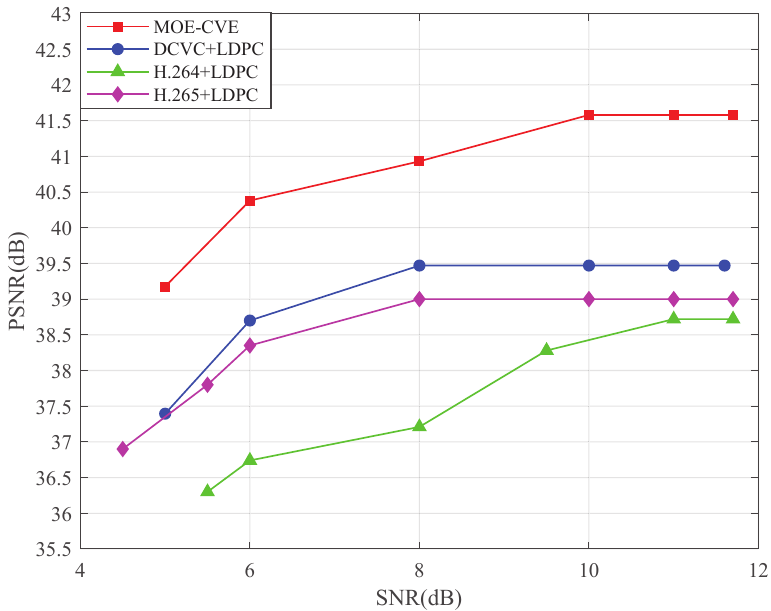}
    \setlength{\abovecaptionskip}{0cm}
    \caption{PSNR of different algorithms under different SNR with CBR=0.015 for H.264/H.265 schemes, and 0.008 for others.}\label{fig:psnr-snr}
    \vspace{-0.7cm}
\end{figure}
Fig.~\ref{fig:psnr-snr} shows how the PSNR changes as the wireless channel SNR varies. 
From Fig.~\ref{fig:psnr-snr}, we see that the PSNR of our method is higher than that of other methods. 
As the SNR increases, the PSNR of both our method and the baselines increase, while our MOE-CVE method increases more fastly. 
This is due to the fact that our method focuses on the major object, and all data amount is used to transmit the semantic feature of the major object, rather than to transmit both the background information. 
From Fig.~\ref{fig:psnr-snr}, we can also observe that the PSNR of the H264+LDPC remains unchanged when SNR is larger than 10~dB. 
When the SNR is greater than 8~dB, the PSNR of our method remains unchanged. 
This demonstrates that our method is more robust against channel noise, especially under low SNR regions. 

\vspace{-0.3cm}
\section{Conclusion}
In this paper, we proposed a MOE-CVE for video semantic communication. 
In particular, we designed a MOE module to extract the major object of the video and a CVE module to encode the semantic information of the major object. 
Simulation results demonstrated that the proposed MOE-CVE scheme can reduce by up to 25\% transmission data while maintaining PSNR and MS-SSIM, compared to the traditional coding schemes. 
Moreover, the proposed MOE-CVE scheme exhibited better robustness to channel impairments than baseline methods. 
\vspace{-0.3cm}

\def\baselinestretch{0.80}
\bibliographystyle{IEEEtran}
\bibliography{CL_edge/ref}

\begin{thebibliography}{10}
\providecommand{\url}[1]{#1}
\csname url@samestyle\endcsname
\providecommand{\newblock}{\relax}
\providecommand{\bibinfo}[2]{#2}
\providecommand{\BIBentrySTDinterwordspacing}{\spaceskip=0pt\relax}
\providecommand{\BIBentryALTinterwordstretchfactor}{4}
\providecommand{\BIBentryALTinterwordspacing}{\spaceskip=\fontdimen2\font plus
\BIBentryALTinterwordstretchfactor\fontdimen3\font minus \fontdimen4\font\relax}
\providecommand{\BIBforeignlanguage}[2]{{%
\expandafter\ifx\csname l@#1\endcsname\relax
\typeout{** WARNING: IEEEtran.bst: No hyphenation pattern has been}%
\typeout{** loaded for the language `#1'. Using the pattern for}%
\typeout{** the default language instead.}%
\else
\language=\csname l@#1\endcsname
\fi
#2}}
\providecommand{\BIBdecl}{\relax}
\BIBdecl

\bibitem{yzh}
Z.~Yang, M.~Chen, Z.~Zhang, and C.~Huang, ``Energy efficient semantic communication over wireless networks with rate splitting,'' \emph{IEEE Journal on Selected Areas in Communications}, May 2023.

\bibitem{videoNumber}
S.~Yang, F.~Li, S.~Trajanovski, R.~Yahyapour, and X.~Fu, ``Recent advances of resource allocation in network function virtualization,'' \emph{IEEE Transactions on Parallel and Distributed Systems}, Feb. 2021.

\bibitem{gwt}
W.~Gong, H.~Tong, S.~Wang, Z.~Yang, X.~He, and C.~Yin, ``Adaptive bitrate video semantic communication over wireless networks,'' \emph{arXiv preprint arXiv:2308.00531}, Aug. 2023.

\bibitem{bandwidthConstrain}
A.~Pratap, R.~Gupta, V.~S.~S. Nadendla, and S.~K. Das, ``Bandwidth-constrained task throughput maximization in {IoT}-enabled {5G} networks,'' \emph{Pervasive and Mobile Computing}, Nov. 2020.

\bibitem{thn}
H.~Tong, Z.~Yang, S.~Wang, Y.~Hu, W.~Saad, and C.~Yin, ``Federated learning based audio semantic communication over wireless networks,'' in \emph{2021 IEEE Global Communications Conference (GLOBECOM)}, Dec. 2021.

\bibitem{semanticSystemExample}
H.~Xie, Z.~Qin, G.~Y. Li, and B.-H. Juang, ``Deep learning enabled semantic communication systems,'' \emph{IEEE Transactions on Signal Processing}, Apr. 2021.

\bibitem{semanticSystemExampleImage}
D.~Huang, X.~Tao, F.~Gao, and J.~Lu, ``Deep learning-based image semantic coding for semantic communications,'' in \emph{Proc. Global Communications Conference (GLOBECOM)}, Dec. 2021.

\bibitem{multimodal}
H.~Xie, Z.~Qin, X.~Tao, and K.~B. Letaief, ``Task-oriented multi-user semantic communications,'' \emph{IEEE Journal on Selected Areas in Communications}, Sept. 2022.

\bibitem{DVC}
G.~Lu, W.~Ouyang, D.~Xu, X.~Zhang, C.~Cai, and Z.~Gao, ``Dvc: An end-to-end deep video compression framework,'' in \emph{Proc. IEEE/CVF Conference on Computer Vision and Pattern Recognition (CVPR)}, Jun. 2019.

\bibitem{SemanticCommunicationsFirst}
P.~Jiang, C.-K. Wen, S.~Jin, and G.~Y. Li, ``Wireless semantic communications for video conferencing,'' \emph{IEEE Journal on Selected Areas in Communications}, Jan. 2023.

\bibitem{ZhangPing}
S.~Wang, J.~Dai, Z.~Liang, K.~Niu, Z.~Si, C.~Dong, X.~Qin, and P.~Zhang, ``Wireless deep video semantic transmission,'' \emph{IEEE Journal on Selected Areas in Communications}, Jan. 2023.

\bibitem{uav}
P.~Si, J.~Zhao, K.-Y. Lam, and Q.~Yang, ``Uav-assisted semantic communication with hybrid action reinforcement learning,'' in \emph{Proc. IEEE Global Communications Conference (GLOBECOM)}, Dec. 2023.

\bibitem{nonlinear}
J.~Dai, S.~Wang, K.~Tan, Z.~Si, X.~Qin, K.~Niu, and P.~Zhang, ``Nonlinear transform source-channel coding for semantic communications,'' \emph{IEEE Journal on Selected Areas in Communications}, Aug. 2022.

\bibitem{MOE}
S.~Lin, L.~Yang, I.~Saleemi, and S.~Sengupta, ``Robust high-resolution video matting with temporal guidance,'' in \emph{Proc. IEEE/CVF Winter Conference on Applications of Computer Vision (WACV)}, Jan. 2022.

\bibitem{loss-moe-lap}
Q.~Hou and F.~Liu, ``Context-aware image matting for simultaneous foreground and alpha estimation,'' in \emph{Proc. IEEE/CVF International Conference on Computer Vision (ICCV)}, Oct. 2019.

\bibitem{opticalEstimation}
A.~Ranjan and M.~J. Black, ``Optical flow estimation using a spatial pyramid network,'' in \emph{Proc. IEEE Conference on Computer Vision and Pattern Recognition (CVPR)}, Aug. 2017.

\bibitem{rANS}
J.~Duda, ``Asymmetric numeral systems: entropy coding combining speed of huffman coding with compression rate of arithmetic coding,'' \emph{arXiv preprint arXiv:1311.2540}, Fan. 2014.

\bibitem{mdvsc}
Z.~Bao, H.~Liang, C.~Dong, C.~Li, X.~Xu, and P.~Zhang, ``Mdvsc--wireless model division video semantic communication,'' \emph{arXiv preprint arXiv:2308.05338}, May. 2023.

\bibitem{MOEdataset}
S.~Lin, A.~Ryabtsev, S.~Sengupta, B.~L. Curless, S.~M. Seitz, and I.~Kemelmacher-Shlizerman, ``Real-time high-resolution background matting,'' in \emph{Proc. the IEEE/CVF Conference on Computer Vision and Pattern Recognition (CVPR)}, Aug. 2021.

\bibitem{DCVCtrain}
T.~Xue, B.~Chen, J.~Wu, D.~Wei, and W.~T. Freeman, ``Video enhancement with task-oriented flow,'' \emph{International Journal of Computer Vision}, Feb. 2019.

\end{thebibliography}

\end{document}